\RequirePackage{ifpdf}
\documentclass[12pt,letterpaper]{JHEPfont}         
\usepackage{amssymb,amsfonts}
\usepackage{float}
\usepackage{units}
\usepackage{amsmath}
\usepackage{amssymb}
\usepackage{graphicx}
\usepackage{color}
\usepackage{esint}
\usepackage{youngtab}

\usepackage{wrapfig}
\usepackage{epsfig}

\def\be{\begin{eqnarray}}
\def\ee{\end{eqnarray}}
\newcommand{\nn}{\nonumber}
\newcommand\para{\paragraph{}}

\newcommand{\eqn}[1]{(\ref{#1})}

\def\Dslash{\,\,{\raise.15ex\hbox{/}\mkern-12mu D}}
\def\Dbarslash{\,\,{\raise.15ex\hbox{/}\mkern-12mu {\bar D}}}
\def\delslash{\,\,{\raise.15ex\hbox{/}\mkern-9mu \partial}}
\def\delbarslash{\,\,{\raise.15ex\hbox{/}\mkern-9mu {\bar\partial}}}
\def\pslash{\,\,{\raise.15ex\hbox{/}\mkern-9mu p}}
\def\calDslash{\,\,{\raise.15ex\hbox{/}\mkern-12mu {\cal D}}}

\newcommand{\bra}{\langle}
\newcommand{\ket}{\rangle}

\def\ket#1{|{#1}\rangle}
\def\bra#1{\langle{#1}|}

\def\lae{\mathrel{\mathop{\smash{\lower .5 ex \hbox{$\stackrel<\sim$}}}}}
\def\lae{\mathrel{\mathop{\smash{\lower .5 ex \hbox{$\stackrel>\sim$}}}}}

\title{\bf  Non-Abelian 3d Bosonization and Quantum Hall States}

\author{   {   \fontencoding{T1}\selectfont {  \DJ or\dj e Radi\v cevi\'c${}^1$, David Tong${}^2$ and Carl Turner${}^2$   }}   \\
${}^1$Stanford Institute for Theoretical Physics and Department of Physics \\
Stanford University, Stanford, CA 94305-4060, USA
 \\
${}^2$Department of Applied Mathematics and Theoretical Physics, \\
University of Cambridge, Cambridge, CB3 OWA, UK \\
{\tt  djordje@stanford.edu, d.tong, c.p.turner@damtp.cam.ac.uk}\\
}

\abstract{Bosonization dualities relate two different Chern-Simons-matter theories, with bosonic matter on one side replaced by fermionic matter on the other. We first describe a more general class of non-Abelian bosonization dualities. We then explore the non-relativistic physics of these theories in  the quantum Hall regime. The bosonic theory lies in a condensed phase and admits vortices which are known to form a non-Abelian quantum Hall state. We ask how this same physics arises in the fermionic theory. We find that a condensed boson corresponds to a fully filled Landau level of fermions, while bosonic vortices map to fermionic holes. We confirm that the  ground state of the two theories is indeed described by the same quantum Hall wavefunction.}

\begin{document}
\pagestyle{plain} \setcounter{page}{1}
\newcounter{bean}
\baselineskip16pt \setcounter{section}{0}

\section{Introduction}

Recent years have seen great progress in our understanding of dualities  in $d=2+1$ dimensional quantum field theories where, for once, we have managed to shrug off the holomorphic comfort blanket of supersymmetry. These developments have arisen from a wonderfully disparate array of topics, including the study of holography, the non-Fermi liquid state of the half-filled Landau level, and the surface physics of topological insulators.

\para
Underlying many of these results is the idea of {\it bosonization}. Roughly speaking, this states that  theories of scalars interacting with $U(N)_k$ Chern-Simons theories are equivalent to theories of fermions interacting with $U(k)_N$ Chern-Simons theories. (More precise statements will be made later in this introduction.) These dualities were originally conjectured in the limit of large $N$ and $k$ \cite{guy1,shiraz,guy2},  motivated in part by their connection to higher spin theories in AdS$_4$ (recently reviewed in \cite{giombi}). They have subsequently been subjected to a  battery of very impressive tests \cite{shiraz2,karthik,shiraz3,sean}.

\para
Versions of these dualities are also believed to hold for finite $N$ and $k$. The first arguments in favour of their existence were given in \cite{shiraz4}, and the first precise dualities were described by Aharony \cite{ofer} by piecing together
evidence from level-rank dualities \cite{levelrank}, known supersymmetric dualities   \cite{givkut,benini,ofer3d4d,parksandrecreation,jain,guy3}, and the map between monopole and baryon operators \cite{radicevic}.

\para
When extrapolated to $N=1$, the dualities imply relationships between Abelian gauge theories, some of  which had been previously proposed  \cite{john}. An example of such a duality equates a theory of bosons, coupled to a Chern-Simons gauge field, to a free fermion. (Closely related conjectures, which differ in some details, have long been a staple of the condensed matter literature --- see, for example,  \cite{polyakov,cfw,shankar1,shankar2,fradhap}.)  Recently it was shown that these Abelian bosonization dualities can be used to derive a whole slew of further dualities  \cite{karchtong,ssww}, including the familiar bosonic particle-vortex duality \cite{peskin,dh}, as well as its more novel fermionic version \cite{son,senthil, ashvin}. The upshot is that there is a web of $d=2+1$ Abelian dualities, with bosonization lying at its heart.

\subsection*{Non-Abelian Bosonization Dualities}

In this paper, our interest lies in the non-Abelian versions of the bosonization dualities. For these, it is a little too quick to say that they relate $U(N)_k$ bosons to $U(k)_N$ fermions since there are subtleties in identifying the levels of the $U(1)$ factors on both sides. These subtleties were largely addressed in \cite{ofer} and, more recently, in \cite{hsin}. Before proceeding, we review these results and provide a slight generalisation.

\para
{\bf Theory A:}  We start by describing the bosonic theory. This consists of $N_f$ scalar fields with quartic couplings,  transforming in the fundamental representation of the gauge group
\be U(N)_{k,\,k'}= \frac{U(1)_{k'N}\times SU(N)_{k}}{{\bf Z}_{N}}\label{upkk}\ee
Here $k$ and $k'N$ denote the levels of the $SU(N)$  and $U(1)$ Chern-Simons terms respectively, so that the action governing the gauge fields is given by
\be {\cal L}_{A} = \frac{k}{4\pi}{\rm Tr}\,\epsilon^{\mu\nu\rho}(a_\mu \partial_\nu a_\rho - \frac{2i}{3} a_\mu a_\nu a_\rho) + \frac{k'N}{4\pi} \epsilon^{\mu\nu\rho} \tilde{a}_\mu\partial_\nu\tilde{a}_\rho \label{lnkk}\ee
with $a$ the $SU(N)$ gauge field and $\tilde{a}$ the $U(1)$ gauge field. (Regularization of each Chern-Simons theory by a small Yang-Mills term is understood throughout.)

\para
The discrete quotient in \eqn{upkk} restricts the allowed values of $k'$ to take the form
\be k'=k+nN\ \ \ \ {\rm with} \ n\in {\bf Z}\nn\ee
A simple way to see this is to construct the $u(N)$-valued gauge field  $a_{u(N)} = a+ \tilde{a}{\bf 1}_N$; the action \eqn{lnkk} becomes a Chern-Simons action for $a_{u(N)}$ at level $k$,  which we denote as $U(N)_k$, together with an Abelian  Chern-Simons action for ${\rm Tr}\,a_{u(N)}$ at level $n$.

\para
The dual of Theory A depends on the choice of Abelian Chern-Simons level $k'$ or, equivalently, on $n$. For $n=0,1$ and $\infty$, the duals were first proposed by Aharony \cite{ofer}. More recently, Hsin and Seiberg described the dual for the choice $n=-1$ \cite{hsin}. Although not explicitly stated by the authors, the techniques of \cite{hsin} allow for a straightforward generalisation\footnote{This generalisation was also noticed by Ofer Aharony and we thank him for extensive discussions on this issue.} to any $n$, which we now describe.

\para
{\bf Theory B:} This consists of $N_f$ fermions, transforming under the fundamental representation of the gauge group $U(k)_{-N+N_f/2}$. The $U(1)\subset U(k)$ gauge field also interacts through  a minimal BF coupling with a further $U(1)_n$ Chern-Simons theory. The resulting action for the gauge fields is
\be {\cal L}_B &=& \frac{-N+N_f/2}{4\pi}\left[  {\rm Tr}\,\epsilon^{\mu\nu\rho}(c_\mu \partial_\nu c_\rho - \frac{2i}{3} c_\mu c_\nu c_\rho)
 + k\, \epsilon^{\mu\nu\rho} \tilde{c}_\mu\partial_\nu\tilde{c}_\rho\right]
\label{correct}\\ &&\ \ \ \  +\ \frac{k}{2\pi}  \epsilon^{\mu\nu\rho} \tilde{c}_\mu\partial_\nu b_\rho + \frac{n}{4\pi}  \epsilon^{\mu\nu\rho} b_\mu\partial_\nu b_\rho
\nn\ee
with $c$ the $SU(k)$ gauge field and  $\tilde{c}, b$ both $U(1)$ gauge fields.

\para
For certain values of $n$, we can integrate out the auxiliary gauge field $b$. These values give the dualities
\be n=\infty: \ \ \ \ \  N_f\ {\rm  scalars\ with}\ SU(N)_k\ \ &\longleftrightarrow&\ \  N_f\ {\rm fermions\ with}\  U(k)_{-N+N_f/2} \nn\\
 n=0: \ \ \  \ \ \  \ N_f\ {\rm  scalars\ with}\ U(N)_k\ \ &\longleftrightarrow&\ \  N_f\ {\rm fermions\ with}\  SU(k)_{-N+N_f/2} \nn\\
n=\pm 1: \ \ \, N_f\ {\rm  scalars\ with}\ U(N)_{k,\,k\pm 1}\ \ &\longleftrightarrow&\ \  N_f\ {\rm fermions\ with}\  U(k)_{-N+N_f/2,\,-N\mp k + N_f/2}
\nn\ee
These are the dualities previously described in \cite{ofer} (for $n=0,1$ and $\infty$) and in \cite{hsin} (for $n=-1$). For general $n$, we cannot integrate out $b$ without generating fractional Chern-Simons levels. In this case, the correct form of the duality is \eqn{correct}.

\subsubsection*{Exploring Quantum Hall States}

The purpose of this paper is to provide evidence for the bosonization dualities described above by studying each theory in the quantum Hall regime. To access this regime, we need to deform both sides of the duality. This is achieved by first  turning on mass deformations so that the theories sit in a gapped phase.   We then
we take the non-relativistic limit.  This involves taking the mass  to infinity while simultaneously turning on a chemical potential which is tuned to the gap. (See, for example, \cite{lmm}  for more details on how to take this limit.)

\para
The retreat to a non-relativistic corner of the theories throws away much of the dynamics that makes bosonization dualities non-trivial. Indeed, here the dualities are souped-up version of  flux attachment, which is used to transmute the statistics of particles in quantum mechanics \cite{frank}. Nonetheless, there remains a lot of interesting physics to extract in this limit and a number of conceptual issues must be understood before we will ultimately find agreement between the two theories.

\para
The full Lagrangians for the bosonic and fermionic non-relativistic theories will be described in Sections \ref{bosonicsec} and \ref{fermionicsec} respectively. In short, they are

\para
{\bf Theory A:} $U(N)_{k,\, k+nN}$ coupled to $N_f$ fundamental scalars.
\para
{\bf Theory B:} $U(k)_{-N+N_f}$ coupled to $N_f$ fundamental fermions and, through a BF coupling, to $U(1)_n$.

\para
Note the shift in the Chern-Simons level of the fermionic theory from $-N+N_f/2$ to $-N+N_f$; this arises because taking the non-relativistic limit involves integrating out the Dirac sea of filled fermionic states.

\para
The dynamics of Theory A is particularly rich in a phase where the gauge symmetry is fully broken so that the theory admits topological vortex solutions. This only occurs when $N_f\geq N$.
 In this paper, we will focus on the specific case $N_f=N$, which is the minimal number of flavours to support such vortices. The two dual theories are then

\para
{\bf Theory A:} $U(N)_{k,\, k+nN}$ coupled to $N_f=N$ fundamental scalars.
\para
{\bf Theory B:} $U(k)_{0}$ coupled to $N_f=N$ fundamental fermions and, through a BF coupling, to $U(1)_n$

\para
We note in passing that there are few concrete tests of the bosonization dualities with $N_f>1$ and, indeed,  it is thought to fail for $N_f$ suitably large \cite{hsin,chiming}. Here we provide a fairly detailed test of the dualities with $N_f=N$.

\para
Our interest in this paper lies in the quantum Hall regime of the two dual theories. As we will see in some detail below, this occurs when the two theories are subjected to a chemical potential for their $U(1)$ factors. We will ultimately find that both theories describe the same quantum Hall states, but the way this arises in the two cases is rather different.

\para
 In Theory A, the emergence of quantum Hall physics involves the condensation of the scalar field and the dynamics of the resulting vortices; this has been studied in some detail in recent (and not so recent) papers  \cite{unknown,carlme,matrix-qhe,matrix-wzw} and will be reviewed in Section \ref{bosonicsec} below. In contrast, in Theory B there is no scalar field to condense. This immediately poses the question: what is the dual of the condensed phase, and what excitations are dual to vortices? We will answer this in Section 3. We will show that the fermions experience an effective background magnetic field, and the dual of the condensed phase is a fully filled Landau level; the vortices are dual to holes in this Landau level.

\section{The Bosonic Theory}\label{bosonicsec}

Theory A consists of $N_f=N$ non-relativistic scalars $\phi_i$, interacting with a $U(N)_{k,\,k'}$ Chern-Simons (CS) gauge field. The complete action is
\be S = \int d^3x\  &&\left[ i\phi_i^\dagger{\cal D}_0 \phi_i  - \,\frac{1}{2m} \vec{\cal D} \phi_i^\dagger\cdot\vec{\cal D} \phi_i
    -\, \frac{\pi}{mk'}(\phi_i^\dagger\phi_i)^2 - \frac{\pi}{mk} (\phi_i^\dagger t^\alpha\phi_i)(\phi^\dagger_jt^\alpha\phi_j)\right] \nn\\
   &&\   +\ \frac{k'N}{4\pi} \epsilon^{\mu\nu\rho} \tilde{a}_\mu\partial_\nu\tilde{a}_\rho + \frac{k}{4\pi}{\rm Tr}\,\epsilon^{\mu\nu\rho}(a_\mu \partial_\nu a_\rho - \frac{2i}{3} a_\mu a_\nu a_\rho) -\mu N \tilde{a}_0\ \ \  \ \ \ \  \ \ \  \label{bosact} \ee
Some comments on conventions:  $i=1,\ldots,N_f=N$ runs over the flavours; the $SU(N)$ generators $t^\alpha$ are in the fundamental representation; $a_\mu$ is the $SU(N)$ gauge field and $\tilde{a}_\mu$ the $U(1)$ gauge field; each scalar $\phi_i$  transforms in the fundamental of $SU(N)$, has charge $1$ under $U(1)$ and has mass $m$. As we mentioned in the introduction, the Chern-Simons level must take the form $k'=k+nN$ with $n\in {\bf Z}$.

%
%

 %
 %


\para
The quartic terms in the first line of \eqn{bosact} are a remnant of similar interactions in the parent, relativistic Theory A which contained Wilson-Fisher scalars. In the non-relativistic context, they give rise to delta-function interactions between particles. The final term in the action is a $U(1)$ charge density $\mu N$. This will prove to be important.

\para
The Gauss law constraints for both Abelian and non-Abelian gauge fields are
\be
\frac{k'N}{2\pi}\,\tilde{f}_{12}=  \phi_{i}^{\dagger}\phi_{i}-\mu N\ \ \ ,\ \ \ \frac{k}{2\pi}\,f_{12}^\alpha =\phi^\dagger_i t^\alpha\phi_i
\label{bosgauss}\ee
There are two translationally invariant ground states which are degenerate in energy:
%
\be  \mbox{\underline{\bf Phase 1:}}\ \ \ \ \ \ \ \  &&  \tilde{f}_{12} = - \frac{2\pi\mu}{k'} \ \ \ , \ \ \ \phi_i=0  \nn \\ \mbox{\underline{\bf Phase 2:}}  \ \ \ \ \ \ \ \ && \tilde{f}_{12} = 0\ \ \ \  ,  \ \ \  \phi_i^\dagger\phi_i =\mu N\nn\ee
%
%
%
In Phase 1, the $U(N)$ gauge group is unbroken and the dynamics includes a Chern-Simons gauge field. This is the quantum Hall phase. However, our interest will initially lie in phase 2, in which the scalars condense and the gauge group is broken. The story we want to explore is how Phase 1 emerges from Phase 2.

\para
We have chosen to work  the smallest number of flavours, $N_f=N$, which can break the gauge symmetry completely. In Phase 2, the scalars pick up expectation values $\phi_{i,a} = \sqrt{\mu}\,\delta_{ia}$ with $a=1,\ldots N$ the gauge index and $i=1,\ldots,N$ the flavour index. The resulting symmetry breaking pattern is
\be U(N)_{\rm gauge}\times SU(N)_{\rm flavour}\ \longrightarrow\ SU(N)_{\rm diag}\label{symmetry}\ee

\subsubsection*{Vortices}

The condensed state admits a new class of excitations: vortices. These are BPS: they are solutions to the  Gauss law constraints \eqn{bosgauss}, together with the first order equation
\be {\cal D}_z\phi_i=0\nn\ee
The single vortex solution has Abelian flux $\int \tilde{f}_{12} = -2\pi/N$. Such fractional flux is allowed because of the ${\bf Z}_N$ quotient in the gauge group \eqn{upkk}.
Inside the vortex, the $\phi$ field decays to zero and Gauss' law \eqn{bosgauss} ensures that the vortices are accompanied by a charge deficit  of $k'$ relative to the condensate.

\para
Interesting things happen when we consider a large number  of vortices together. The resulting physics was studied in some detail in \cite{matrix-qhe,matrix-wzw}, following earlier work on the Abelian theory \cite{unknown,carlme}.  Here we summarise the main results.

\para
The BPS nature of the vortices means that there is no unique classical solution; in particular, the vortices can be placed anywhere on the plane. It is simple to rectify this by adding a harmonic trap which forces the vortices towards the origin. (Such a trap is most easily constructed by taking it proportional to the angular momentum of the vortex configuration.) In the presence of this  trap, there is a unique minimum energy vortex configuration which consists of a large, circular droplet, inside of which
$\phi_i=0$ and $\tilde{f}_{12} = -2\pi \mu / k'$.
The total flux carried by $M$ vortices is simply $-2\pi M/N$. Equating this to the flux  $[-2\pi \mu/k']\pi R^2$, we learn that the area of the droplet containing $M$ vortices is
\be \pi R^2 \approx \frac{k'M}{\mu N}
\label{vortexarea}\ee
The upshot of this argument is that adding a macroscopically large number, $M$, of vortices creates a macroscopically large region of space in which the gauge symmetry is unbroken. In other words, we have succeeded in constructing a finite region of Phase 1 (the quantum Hall phase) that sits inside Phase 2.


\para
The advantage of  this construction is that the vortices also give us a handle on microscopic aspects of the  quantum Hall state. In particular, by quantising the low-energy dynamics of the vortices, we can reconstruct various properties of the quantum Hall state. First, let's build some expectations. One key fact is that as the  vortices move, they experience a background magnetic field. This follows from a simple duality argument: the term $\mu N\tilde{a}_0$ in the action \eqn{bosact} is a background charge density for electric states, but acts like a background magnetic field $B=2\pi\mu$ for magnetic states. On general grounds, we expect that the density of states in the lowest Landau level is given by $B/2\pi = \mu$. Yet, from   \eqn{vortexarea}, we see that the density of vortices sitting in our droplet is $\rho_v = \mu N/k'$. This suggests that the quantum Hall state of vortices has filling fraction
\be \nu = \frac{\rho_v}{B/2\pi} = \frac{N}{k'} \label{filling}\ee
The next step is to understand the quantum Hall wavefunctions which describe this state. Here there are two possible methods: one direct, one indirect:
\begin{itemize}
\item The direct method is to construct the quantum mechanics of $M$ vortices and solve for its ground state wavefunction. This involves solving a complicated many-body system and, in general, is  not easy. Nonetheless, as we review below, progress can be made in the special case of $k'=k+N$ (or $n=1$).
\item For a more indirect method, recall that the gauge group is unbroken inside the droplet of vortices, but broken outside. This means that the low-energy dynamics  includes a $U(N)_{k,\,k'}$ Chern-Simons theory which, on the edge of the droplet, induces a $U(N)_{k,\,k'}$ WZW conformal field theory. Now we use an insight due to Moore and Read \cite{mr} sometimes known as the {\it bulk-boundary correspondence}. (It can be thought of as a baby version of de Sitter holography.) This says that  the bulk quantum Hall wavefunction can be identified with a suitable correlation function in the boundary conformal field theory. (A review of the bulk-boundary correspondence  applied to  quantum Hall physics can be found in the lecture notes \cite{me}.)
\end{itemize}
We now review how we can construct the quantum Hall states using both of these methods.

\subsubsection*{The Indirect Method: Conformal Field Theory}

We start with the indirect method in which the wavefunction is identified with a suitable correlation function of the $U(N)_{k,\,k'}$ WZW conformal field theory.

\para
For $N=1$, the WZW model is simply a  compact boson and the resulting wavefunctions are the Laughlin states \cite{mr}. For $N>1$, the appropriate correlation functions were first computed by Blok and Wen and give rise to non-Abelian quantum Hall states.  A slightly different presentation of these wavefunctions was offered in \cite{matrix-qhe} and this is the notation we use here.

\para
Let us first think about the kind of wavefunction that we expect.  It is simple to check that a single vortex transforms in the $k^{\rm th}$ symmetric representation of the $SU(N)_{\rm diag}$ symmetry \eqn{symmetry}. This means that the vortex carries an internal ``spin" degree of freedom; the wavefunction will depend on both the position $z$ and the spin $\sigma$ of each vortex.

\para
The $SU(N)$ quantum numbers are sufficient to identify the boundary operator ${\cal O}_R$ that corresponds to the vortex: it is the primary operator with $R$ the $k^{\rm th}$ symmetric representation. Roughly speaking, we then identify the bulk wavefunction as
\be \Psi(z,\sigma)  \sim \langle {\cal O}_R(z_1)\ldots{\cal O}_R(z_M)\rangle\nn\ee
(A more precise statement involves a careful treatment of the $U(1)\subset U(k)$ part of the WZW model; details can be found in \cite{matrix-qhe}.)

\para
We describe the wavefunction when the number of vortices, $M$, is divisible by $N$. Here things are somewhat simpler as  the wavefunction turns out to be an $SU(N)$ singlet. To proceed, it is useful to attach an internal state $|\sigma_a\rangle$ to each vortex, with $\sigma_a\in \{1,\ldots,N\}$. This is slightly unnatural because, as we mentioned above, the vortices transform in the $k^{\rm th}$ symmetric product of $SU(N)$ rather than the fundamental. However, in the wavefunction there will be $k$ states $\ket{\sigma_{a_m}}$ per vortex, suitably symmetrised, filling out this representation.
We  define the {\it baryon} to be a combination of $N$ vortices, with auxiliary spins arranged to form a singlet: $B_{a_1\ldots a_N} = \epsilon^{\sigma_{a_1}\ldots \sigma_{a_N}} \ket{\sigma_{a_1}}\ldots \ket{\sigma_{a_N}}$. The correlation function in the CFT gives the bulk wavefunction in the form
\be \Psi(z,\sigma) = \prod_{a<b}^{M} (z_a-z_b)^n\, {\rm Sym}\left[ \Phi^k(z,\sigma)\right] e^{-2\pi\mu \sum_a|z_a|^2/4} \label{final}\ee
where
\be \Phi(z,\sigma) &=& \epsilon_{a_1\ldots a_{M}} (z_{a_1} \ldots z_{a_N})^0 (z_{a_{N+1}} \ldots z_{a_{2N}})^1 \ldots (z_{a_{M-N+1}}\ldots z_{a_{M}})^{M/N-1} \nn\\
&& \ \ \ \ \ \ \times\ B_{a_1\ldots a_N}B_{a_{N+1}\ldots a_{2N}}\ldots B_{a_{M-N+1}\ldots a_{M}}\label{Phi}\ee
and the symbol $\rm Sym[\ldots]$ projects onto the symmetrised product of spin states, ensuring  that each particle transforms in the  $k^{\rm th}$ symmetric representation of $SU(N)$.

\para
The wavefunctions \eqn{final} are the Blok-Wen states.  They  have the anticipated filling fraction \eqn{filling}. They are an example of
an  $N$-clustered state, meaning that the wavefunction vanishes only if $N+1$ or more particles coincide. For $N=2$ and $k=2$, the wavefunction is a spin-singlet generalisation of the well-known Moore-Read state \cite{mr}. For $N>2$ and $k=2$, it is a spin-singlet generalisation of the Read-Rezayi states \cite{rr}.

\subsubsection*{The Direct Method: Vortex Matrix Model}

The method described above requires us to invoke the somewhat magical correspondence between boundary correlation functions and bulk wavefunctions. A much more direct approach is as follows: determine the interactions between vortices and then solve for the ground state wavefunction. Both of these steps are difficult and in general there is no reason to believe that this is any easier  than other many-body problems.  Nonetheless, progress can be made in the special case of
\be k'=k+N\ \ \ \ \ ({\rm or}\ n=1)\nn\ee
In this case, one can construct a description of the vortex dynamics in terms of a $U(M)$ matrix model. This was studied in detail in  \cite{unknown,carlme,matrix-qhe}\footnote{A warning on  notation: in the matrix model papers \cite{carlme,matrix-qhe} we described $N$ vortices in a $U(p)$ Chern-Simons theory by a  $U(N)$ matrix quantum mechanics. This, of course, differs from the use of these variables in the present paper where we have instead opted for consistency with the bosonization literature.}.
 The matrix model turns out to be  solvable and allows us to determine in  the properties of the vortex ground state as well as the spectrum of excited states. Here we describe only the main results

\begin{itemize}
\item When $N=1$, we have an Abelian quantum Hall state. The vortex dynamics was shown in \cite{unknown,carlme} to be described by a  matrix model previously studied by Polychronakos \cite{alexios} (who, in turn, was inspired by \cite{lenny}).  The ground state of this matrix model is known to coincide  (asymptotically) with the Laughlin wavefunction \cite{alexios,hellvram,ks}.
\item
For $N>1$, the vortices carry an internal spin which, as we mentioned above, transforms in the  $k^{\rm th}$ symmetric representation of the $SU(N)_{\rm diag}$ symmetry. If we place $M$ vortices in a harmonic trap, then the representation of the resulting ground state depends on
the value of $M$ mod $N$. Writing $M=m$ mod $N$, the configuration of vortices transforms in the
in the $k^{\rm th}$ symmetrisation of the $m^{\rm th}$ antisymmetric representation. In terms of Young diagrams, this is
\be  m\left\{ \begin{array}{c} \\ \\ \end{array}\right. \!\!\!\overbrace{\raisebox{-3.1ex}{\yng(5,5,5)}}^{k}
\label{young}\ee
In particular, when $M$ is divisible by $N$ the ground state is a singlet under $SU(N)$.

\item For $N>1$, the ground state of the matrix model coincides with the Blok-Wen states \eqn{final}.

\item Finally, we can relate this discussion to the indirect method described above. The excitations of the vortex configuration are chiral modes,  living on the edge of the droplet. In the large $N$ limit, the dynamics of these excitations can be shown to coincide with those of the $U(N)_{k,\,k'}$ WZW conformal field theory \cite{matrix-wzw}.
\end{itemize}

\section{The Fermionic Theory}\label{fermionicsec}

Now we turn to the Theory B. Our task is to reproduce the properties of vortices described above  in terms of fermions. The theory consists of
 $N_f=N$ non-relativistic fermions $\psi_i$. These   interact with a  $U(k)_{0}$ gauge field;  we denote the $SU(k)$ part as $c$ and the $U(1)\subset U(k)$ part as $\tilde{c}$. As described in the introduction, this is subsequently coupled to a further $U(1)_n$ gauge field, $b$. The full action is
\be S = \int d^3x\ && \left[ i\psi_i^\dagger {\cal D}_0 \psi_i  - \,\frac{1}{2m} \vec{\cal D} \psi_i^\dagger\cdot\vec{\cal D} \psi_i
    -\psi_i^\dagger G\psi_i\right]  \nn\\ &&\ \ \ \ \ \ \ + \ \frac{k}{2\pi}  \epsilon^{\mu\nu\rho} \tilde{c}_\mu\partial_\nu b_\rho + \frac{n}{4\pi}  \epsilon^{\mu\nu\rho} b_\mu\partial_\nu b_\rho -\frac{\mu k }{n} \tilde{c}_0  \label{fermiact} \ee
%
%
%
 The third term in the action couples the fermions to the background magnetic field, $G=g_{12} + \tilde{g}_{12}{\bf 1}_k$, where  $g=dc - i[c,c]$ and $\tilde{g} = d\tilde{c}$ are the non-Abelian and Abelian field strengths respectively. This term arises from the non-relativistic limit of the Dirac equation.

\para
Note that the duality maps the chemical potential $\mu N$ of Theory $A$ into a chemical potential $\mu k/n$ of Theory B. This map can be explicitly checked (at least in the Abelian case) using the techniques of \cite{karchtong,ssww}; for non-Abelian gauge groups considered here,  the map between chemical potentials includes a rescaling by the rank of the gauge group. As an alternative, one can change the term in \eqref{fermiact} for a chemical potential for $b$; in this case it takes the simpler form $-\mu k \tilde{c}_0/n \to +\mu b_0$. The physics which follows is identical. This allows a clearer extension to $n=0$.

\para
Our task is to reproduce the quantum Hall physics found in the bosonic theory. The essence of the problem becomes immediately apparent if we look at the Gauss' law constraints. Because the $SU(k)$ Chern-Simons level is vanishing, the dynamics of the non-Abelian field $c$ is solely governed by the Yang-Mills regulator whose coupling is taken to be large; thus, this gauge theory is confined and only $SU(k)$ singlets are allowed. In contrast, the Abelian Gauss' law arising from $\tilde{c}$ and $b$ read
\be
\psi_{i}^{\dagger}\psi_{i}-\frac{\mu k}{n} + \frac{k}{2\pi}db &=&0 \label{newgauss}\\
\frac{k}{2\pi} d\tilde{c} + \frac{n}{2\pi} db &=& 0 \nn\ee
Now we see the difficulty. There is only one obvious, translationally invariant  solution, given by $db = -(k/n) d\tilde{c}$ and
\be \mbox{\underline{\bf Phase 1$^\prime$:}}\ \ \ \ \ \ \ \  d\tilde{c} = \tilde{g}_{12} =  -\frac{2\pi\mu}{k}\ \ \ ,\ \ \ \langle\psi_i^\dagger\psi_i\rangle=0\nn\ee
This provides the dual to Phase 1 of the bosonic theory. However, life is more difficult if we want to write down the dual of Phase 2 in the bosonic theory
because we cannot simply condense the fermions to saturate the background charge.  How, then, to construct Phase 2?


\para
To do this, we work self-consistently. Suppose there is a constant, background Abelian field with strength $\tilde{g}_{12}$. The fermionic excitations then form Landau levels. However, crucially, the presence of the $\psi^\dagger_i \tilde{g}_{12}\psi_i$ term in the action \eqn{fermiact} means that the lowest Landau level costs zero energy. (This is a familiar fact for relativistic fermions, and the direct coupling to the field strength arises because \eqn{fermiact} is the non-relativistic limit of a relativistic theory.) This means that there is a second, translationally invariant ground state in which the lowest Landau level is fully filled. The density of states in a Landau level is $|\tilde{g}_{12}|/2\pi$ and, including both flavour and colour degrees of freedom, there are $kN$ different fermions which we can excite. Hence the fully filled lowest Landau level has $\langle \psi_i^\dagger\psi_i\rangle = kN|\tilde{g}_{12}|/2\pi$. The self-consistent solution to \eqn{newgauss} is then
\be  \mbox{\underline{\bf Phase 2$^\prime$:}}\ \ \ \ \ \ \ \  \tilde{g}_{12} =  -\frac{2\pi\mu}{k'}\ \ \ ,\ \ \ \langle\psi_i^\dagger\psi_i\rangle=\frac{\mu k N}{k'}\nn\ee
where $k'=k+nN$. We claim that this phase is dual to Phase 2 of Theory A.\footnote{One could also consider such self-consistent solutions for bosons. In this language, the condensed Phase 2 for bosons corresponds to filling the lowest Landau level an infinite number of times, a luxury not available for fermions. Filling a finite number of times would appear to correspond to a fractionally filled Landau level for the fermions; it would be interesting to explore this connection further.}

\subsubsection*{Holes as Vortices}

Our next task is to understand the excitations above Phase $2'$. These are the dual to the vortices in Theory A.  Since all physical states must be $SU(k)$ singlets, the lowest energy excitations are baryonic holes in the lowest Landau level.  In the absence of a trap, these cost zero energy and are created by operators
\be
H_{i_{1}\ldots i_{k}}(\vec{x})=\epsilon^{m_{1}\ldots m_{k}}\psi_{i_{1}m_{1}}(\vec{x})\cdots\psi_{i_{k}m_{k}}(\vec{x})
\label{hole}\ee
where the colour indices range from $m=1,\ldots,k$ and the flavour indices from $i=1,\ldots,N$.
Gauss' law \eqn{newgauss} ensures that each hole is accompanied by a flux $db = -2\pi$ and $\tilde{g}_{12} = 2\pi n/k$.
 We will now show that these holes share the same properties as the vortices in Theory A.

\para
Theory B has an $SU(N)$ flavour symmetry. In Phase 2$'$, this should be identified with the $SU(N)_{\rm diag}$ symmetry \eqn{symmetry} of Theory A.
Since the fermions in \eqn{hole} are anti-commuting, the hole operators $H_{i_1\ldots i_k}$ must transform in the $k^{\rm th}$ symmetric representation of $SU(N)$. This coincides with the transformation of a single vortex in Theory A.

\para
What happens as we introduce more and more baryonic holes? Clearly, we  start to construct a region that takes us back to Phase 1$'$. Just as it was useful to understand Phase 1 of the bosonic theory through the lens of the vortices, here we would like to understand Phase $1'$ through the lens of the holes. The first step is to notice that the holes feel as if they are moving in a background magnetic field. This is because they carry flux $\tilde{g}_{12} = 2\pi n/k$ and, by the same kind of duality argument we used in Section \ref{bosonicsec}, the $(\mu k/n) \tilde{c}_0$ term in the action mimics a magnetic field for any magnetic excitation. The strength of this effective magnetic field is $B = 2\pi\mu$.

\para
Meanwhile, the maximum density of holes is $\rho_h = \langle\psi_i^\dagger\psi_i\rangle/k = \mu N/k'$, because each hole consists of $k$ $\psi$ excitations. This means that the holes can be  packed at filling fraction
\be \nu = \frac{\rho_h}{B/2\pi} = \frac{N}{k'}\nn\ee
This coincides with the filling fraction of vortices \eqn{filling} that we saw in Theory A.

\para
The mapping of quantum numbers and density provides good evidence that  non-Abelian vortices map to holes in the lowest Landau level. The BPS nature of the vortices is associated to vanishing energy of states in the lowest Landau level.


\para
Our next task is to construct wavefunctions for these states. Since the holes created by $\psi_{ia}$ experience a background magnetic field, wavefunctions for a  single-hole are just the familiar lowest Landau level states. In symmetric gauge, the quantum fields can be expanded in angular momentum modes as
\be \psi_{im}(z,\bar{z}) = \sum_{q=0}^\infty z^q\,e^{-B |z|^2/4} \ \chi^q_{im}\label{expand}\ee
%
%
%
where $\chi^q_{im}$ is the creation operator for a fermion, labelled by $i$ and $m$, in the $q^{\rm th}$ angular momentum state of the lowest Landau level.

\para
We now look at states with $N$ holes. This  is trickier as we should take into account the interaction between holes. We will proceed by neglecting this. Partial justification comes from the fact that the $SU(k)$ gauge interactions are strongest and we have already taken these into account in forming the baryonic holes. Nonetheless, one may expect some residual short range interactions which we do not have control over. The fact that ultimately the ground state is gapped (and the  agreement with the dual description) suggests that this is valid.

\para
 To provide an energetic distinction between different hole excitations, we introduce a harmonic trap. As in Theory A, it is simplest to take the trap to be proportional to the angular momentum $q$ of the holes, with the convention that Phase 2$'$ has vanishing energy.
 For each spatial wavefunction, we have $Nk$ fermionic states $\psi_{im}$. Each hole is constructed from $k$ of these states. This means that the first $N$ holes sit in the lowest, $q=0$, state; the next $N$ holes sit in the $q=1$ state, and so on.

\para
What representation of $SU(N)$ does the resulting ground state sit in? To see this, note that we can equally well write the single hole creation operator  \eqn{hole} as
\be H_{i_1,\ldots,i_k} = {\rm Sym}_i[\psi_{i_1,1}\ldots\psi_{i_k,\,k}]\nn\ee
where the symmetrisation is over all flavour indices. Now consider the product over two, spatially coincident holes,  $H_{i_1,\ldots,i_k}H_{j_1,\ldots,j_k} = {\rm Sym}_{i,j}[\psi_{i_1,1}\ldots\psi_{i_k,\,k}\psi_{j_1,1}\ldots\psi_{j_k,\,k}]$ where we symmetrise independently over $i$ indices and over $j$ indices. Clearly this state is anti-symmetric under exchange of each pair, such as $(i_1,j_1)$. The upshot is that this state transforms in the $k^{\rm th}$ symmetrisation of the anti-symmetric representation or, in terms of Young diagrams,
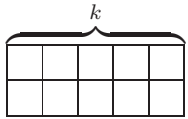
\be   \!\!\!\overbrace{\raisebox{-3.1ex}{\yng(5,5)}}^{k} \nn\ee
By the same argument, we see that the ground state of $M=m$ mod $N$ holes transforms in the same representation \eqn{young} as the ground state of vortices.

\para
Before writing down the many-hole wavefunction, there is one final thing we should remember. The holes are composite fermions/bosons; they have charge $k$ and flux $2\pi n/k$. This means that when one hole circles another, it picks up a $2\pi n$ phase. To reflect this, we should include the factor  $\prod (z_a-z_b)^n$ in the wavefunction.

\para
We've now described all the ingredients which go into constructing the wavefunction for $M$ holes.  The only remaining difficulty is notational. For simplicity, we take $M$ divisible by $N$. Each hole, $a=1,\ldots,M$, has an associated $SU(N)$ spin ${\cal H}_a$ which lies in the $k^{\rm th}$ symmetric representation of $SU(N)$
\be {\cal H}_a(\vec{x}) = (H_a)_{i_1\ldots i_k}(\vec{x})\, \ket{\sigma_{i_1}}\ldots\ket{\sigma_{i_k}}\nn\ee
where, as for the vortices, $\ket{\sigma}\in\{ 1,\ldots,N\}$. The wavefunction is then given by the overlap
\be \Psi(z,\sigma) = \prod_{a<b}^M(z_a-z_b)^n \ \bra{\hspace{0.08em}{\rm LLL}\hspace{0.08em}}\, {\cal H}_{a_1}^\dagger(z_1,\bar{z}_N)\ldots {\cal H}_{a_N}^\dagger(z_N,\bar{z}_N)\,\ket{M}\nn\ee
where $\bra{\hspace{0.08em}{\rm LLL}\hspace{0.08em}}$ is the ground state for Phase $2'$, while $\ket{M}$ is the state with the $M$ holes removed in successive lowest angular momentum modes. To construct the explicit wavefunction now involves only Wick contractions of the creation operators $\chi^q_{im}$ which appear in \eqn{expand}. Despite its simplicity, this step  is a little fiddly.  It is easiest to focus on a specific colour index, say $m=1$. One can check that the resulting terms in the wavefunction are precisely those that appear in $\Phi(z,\sigma)$ defined in \eqn{Phi}. Repeating this for each $m=1,\ldots,k$, we find  the Blok-Wen wavefunction \eqn{final}, where the symmetrization naturally occurs for the reasons described above.

\subsubsection*{Level Rank Duality}

Comparing the construction of the wavefunction for holes and vortices, we see that there is an interesting interplay the roles played by $SU(k)$ and $SU(N)$ on the two sides of the duality. This is the essence of level-rank duality. In this section, we review some representation theory which highlights this connection.

\para
In building the hole wavefunctions, we find that each state in the lowest Landau level comes in $Nk$ varieties, each associated to a fermionic annihilation operator
 $\psi_{i,m}$ with $i=1,\ldots,N$ and $m=1,\ldots,k$. These states naturally carry a representation of $u(Nk)_1$. This then has a decomposition into
\be u(1)_{Nk} \times su(k)_N \times su(N)_k \subset u(Nk)_1\label{decomp}\ee
The first factor, $u(1)_{Nk}$ simply counts the number of excited fermions. The second and third factors correspond to our gauge and flavour groups respectively. (The levels arise because there is a truncation on the dimension of each representation, which follows simply from the fact that we have a finite number of Grassmann operators to play with.) Gauge invariance means that we want to restrict to $SU(k)$  singlet. The question we would like to ask is: which $SU(N)$ representations then emerge?

\para
The general decomposition \eqn{decomp} has been well studied,  not least because of the important role it plays in level-rank duality.  We label representations under the left-hand side using triplets ($q,R,\tilde{R}$), where $q$ is the number of excited fermions and $R$ and $\tilde{R}$ denote the Young diagrams for the representations of $su(k)_N$ and $su(N)_k$ respectively. Suppose that the representation $R$ appears on the left-hand side: then it is accompanied by $\tilde{R} = R^T$, or its orbit under outer automorphisms. Let us first explain what this means.

\para
The {\it outer automorphism group} of $SU(N)_k$ is ${\bf Z}_N$. It is generated by the basic outer automorphism operator $\sigma$ which obeys $\sigma^N=1$. This has an action on representations which can be nicely explained using Young diagrams. We start with a given Young diagram  $\tilde{R}$.  Then $\sigma(\tilde{R})$ is a second Young diagram which we construct using the following procedure: first, add a row of length $k$ to the top of $\tilde{R}$; next
remove any columns of length $N$ to obtain a suitably reduced Young diagram. One may easily verify that this procedure gives $\sigma^N(\tilde{R}) = \tilde{R}$ for any $\tilde{R}$.

\para
The upshot of this is that the only representations of $ u(1)_{Nk} \times su(k)_N \times su(N)_k$ that can appear are
$\big(|R| + mk\,(\mathrm{mod}\, kN),R,\sigma^m(R^T)\big)$, with $m=0,1,\ldots,N-1$.
Here $R^T$ denotes the transpose of the Young diagram $R$, and $|R|$ is the number of boxes it contains.

\para
For us, the above construction is particularly simple because we are interested in the singlet representation $R$. These have $|R|=0$, and $R^T$ is the  singlet representation of $su(p)$. Under the action of outer automorphisms, the singlet representation is mapped into representations which contain $M$ complete rows of $k$ boxes, with $u(1)_{Nk}$ charge $Mk$. This means that the operators $H^M$, with $M<N$, transform in the representation \eqn{young} which we saw for vortices in Theory A.

\para
The discussion above was restricted to $M<N$ baryonic holes. Each spatially distinct state in the lowest Landau level has $Nk$ fermionic states. This means that if we
 remove $N$ baryons then we empty one spatial bucket, leaving the state a singlet once more. Then we must begin again from
the next bucket, and the process repeats. So for $M$ baryonic holes, the representation is again given by \eqn{young}, where $M=m$ mod $N$. This again matches the representation theory of the vortices.

\section{Discussion}

There are a bewildering number of descriptions of quantum Hall states. Many of these are related by dualities, including the kinds of bosonization dualities described in this paper. Here we try to place our results within this wider context.

\para
The original effective field theory for the Laughlin state is due to Zhang, Hansson and Kivelson \cite{zhk}. It consists of an Abelian Chern-Simons theory with non-integer level set by the filling fraction. The Chern-Simons field is coupled to non-relativistic scalars which, through the process of flux attachment, become  the electrons of the system. An alternative description was offered by Lopez and Fradkin, \cite{lopezfrad} which again consists of an Abelian Chern-Simons field at non-integer level, this time coupled to fermions. The equivalence of these two descriptions for the long distance physics can be viewed as a simple example of 3d bosonization, albeit restricted to the non-relativistic regime of quantum mechanics.

\para
The fact that the Chern-Simons level in \cite{zhk,lopezfrad} is fractional means that these theories miss aspects of the physics related to topological order. This was rectified in the work of Wen and Zee \cite{wenzee}, who presented an effective description of quantum Hall states in terms of Abelian Chern-Simons theories with integer-valued levels. These are related to the earlier papers through a kind of particle-vortex duality. In particular, the vortices now play the role of the electrons in the system. The gauge fields are coupled to scalars whose excitations describe the quasi-holes with  anyonic statistics.

 \para
 To our knowledge, a fermionic version of the Wen-Zee class of theories has not previously been constructed. This is what the bosonization duality achieves. For example,  the results of Section 3 tell us that the Laughlin state at filling fraction $\nu = 1/(k+1)$ is described by a $U(k)_{0,-k} \cong [U(1)_{-k^2} \times SU(k)_0]/{\bf Z}_k$ Chern-Simons theory coupled to just a single species of fermion. This viewpoint appears to be closely related to the partonic construction of \cite{jkjain,parton}.

\para
The bosonic ``Theory A" that we have described in Section 2 should be viewed in the same spirit as the Wen-Zee theories, with the obvious exception that it is a non-Abelian gauge theory. It is a $U(N)_{k,\,k'}$ Chern-Simons theory whose vortices are to be thought of  as the ``electrons", now endowed with internal spin degrees of freedom. The resulting quantum Hall states were previously introduced by Blok and Wen. The bosonization duality now tells us that the duals of these non-Abelian states can be constructed by considering $SU(k)$ singlets, coupled to further Abelian gauge fields. This is reminiscent of the partonic description of these states previously presented in
 in \cite{wen,bw}.

\para
Finally, it would be interesting to understand to what extent the bosonization dualities relating \eqn{lnkk} and \eqn{correct} underlie more general non-Abelian dualities in $d=2+1$ dimensions. For example, are they related to other approaches such as \cite{murugan,nunez}? Can they be used as building blocks to derive non-Abelian particle-vortex dualities, or their supersymmetric counterparts constructed in \cite{is,hananywitten}?

\section*{Acknowledgements}

We thank   Ofer Aharony for sharing his insights on the general $k'$ dualities described in the introduction. We are also  grateful to Po-Shen Hsin, Chao-Ming Jian, Andreas Karch,  R.~Loganayagam, Shiraz Minwalla, Nathan Seiberg, Stephen Shenker and Spenta Wadia for useful conversations on various aspects of this work. DT thanks Stanford SITP for their kind hospitality while this work was initiated.
DT and CT are supported  by the European Research Council under the European Union's Seventh Framework Programme (FP7/2007-2013), ERC grant agreement STG 279943, ``Strongly Coupled Systems", and CT is also supported by an STFC studentship. {   \fontencoding{T1}\selectfont {\DJ R}} is supported by a Stanford Graduate Fellowship and thanks the Perimeter Institute for hospitality while this work was being finished. Research at Perimeter Institute is supported by the Government of Canada through Industry Canada and by the Province of Ontario through the Ministry of Economic Development \& Innovation.

\end{document}